\documentclass[twocolumn,english,aps,pra,superscriptaddress,showpacs,tightenlines]{revtex4-1}
\usepackage{amssymb}
\usepackage{amsmath}
\usepackage{amsfonts}
\usepackage{graphicx}
\usepackage{epsfig}
\usepackage{color}
\usepackage{txfonts}

\begin{document}

\title{Nonreciprocal Single-Photon Band Structure in a Coupled-Spinning-Resonator chain}
\author{Jing \surname{Li}}
\affiliation{Key Laboratory of Low-Dimension Quantum Structures and Quantum Control of Ministry of Education, Key Laboratory for Matter Microstructure and Function of Hunan Province, Synergetic Innovation Center for Quantum Effects and Applications, Xiangjiang-Laboratory and Department of Physics, Hunan Normal University, Changsha 410081, China}
\affiliation{Institute of Interdisciplinary Studies, Hunan Normal University, Changsha, 410081, China
}
\author{Ya \surname{Yang}}
\affiliation{School of Physics and Chemistry, Hunan First Normal University, Changsha 410205, China
}
\author{Xun-Wei \surname{Xu}}
\affiliation{Key Laboratory of Low-Dimension Quantum Structures and Quantum Control of Ministry of Education, Key Laboratory for Matter Microstructure and Function of Hunan Province, Synergetic Innovation Center for Quantum Effects and Applications, Xiangjiang-Laboratory and Department of Physics, Hunan Normal University, Changsha 410081, China}
\affiliation{Institute of Interdisciplinary Studies, Hunan Normal University, Changsha, 410081, China
}
\author{Jing \surname{Lu}}
\affiliation{Key Laboratory of Low-Dimension Quantum Structures and Quantum Control of Ministry of Education, Key Laboratory for Matter Microstructure and Function of Hunan Province, Synergetic Innovation Center for Quantum Effects and Applications, Xiangjiang-Laboratory and Department of Physics, Hunan Normal University, Changsha 410081, China}
\affiliation{Institute of Interdisciplinary Studies, Hunan Normal University, Changsha, 410081, China
}
\author{Hui \surname{Jing}}
\affiliation{Key Laboratory of Low-Dimension Quantum Structures and Quantum Control of Ministry of Education, Key Laboratory for Matter Microstructure and Function of Hunan Province, Synergetic Innovation Center for Quantum Effects and Applications, Xiangjiang-Laboratory and Department of Physics, Hunan Normal University, Changsha 410081, China}
\affiliation{Institute of Interdisciplinary Studies, Hunan Normal University, Changsha, 410081, China
}
\author{Lan \surname{Zhou} }
\email{zhoulan@hunnu.edu.cn}

\affiliation{Key Laboratory of Low-Dimension Quantum Structures and Quantum Control of Ministry of Education, Key Laboratory for Matter Microstructure and Function of Hunan Province, Synergetic Innovation Center for Quantum Effects and Applications, Xiangjiang-Laboratory and Department of Physics, Hunan Normal University, Changsha 410081, China}
\affiliation{Institute of Interdisciplinary Studies, Hunan Normal University, Changsha, 410081, China
}

\begin{abstract}
We analyze the single-photon band structure and the transport of a single photon in a one-dimensional coupled-spinning-resonator chain. The time-reversal symmetry of the resonators chain is broken by the spinning of the resonators, instead of external or synthetic magnetic field. Two nonreciprocal single-photon band gaps can be obtained in the coupled-spinning-resonator chain, whose width depends on the angular velocity of the spinning resonator.
Based on the nonreciprocal band gaps, we can implement a single photon circulator at multiple frequency windows, and the direction of photon cycling is opposite for different band gaps. In addition, reciprocal single-photon band structures can also be realized in the coupled-spinning-resonator chain when all resonators rotate in the same direction with equal angular velocity. Our work open a new route to achieve, manipulate, and switch nonreciprocal or reciprocal single-photon band structures, and provides new opportunities to realize novel single-photon devices.
\end{abstract}

\date{\today}
\maketitle


\section{Introduction}

In optical systems, optical nonreciprocity plays a key role in guiding
optical flow and protecting optical elements from backscattered light. In
the fields of classical\cite{2017NaPho..11..441S,Feldmann2021ParallelCP} and
quantum computing\cite{2016Sci...354..847S,Moody_2022}, communication\cite%
{Roelkens2010IIIVsiliconPF,Rizzo2023MassivelySK}, and sensing\cite%
{PhysRevA.103.042418,2019NatMa..18..783O,
PhysRevLett.112.203901,PhysRevA.96.033842,Zhao224}, increasingly complex photonic
integrated circuits require controllable, on-chip nonreciprocal forms. With
the continuous evolution of photonic integrated circuits towards greater
intricacy, the demand for customizable on-chip non-reciprocal
functionalities is more pronounced than ever. One of the most basic
requirements for realizing nonreciprocal transmission in optical systems is
to break the symmetry of time inversion\cite{1964ApOpt...3..544A}.
Traditionally, nonreciprocal devices leaned on ferromagnetic materials,
harnessing the Faraday rotation effect to disturb time reversal symmetry.
Nonetheless, this approach grapples with intricate procedures and the
necessity for potent magnetic fields, posing challenges for achieving
on-chip integration of optical nonreciprocal devices. Over recent years,
researchers have unveiled a myriad of innovative theoretical frameworks and
experimental methodologies aimed at circumventing the stringent limitations
imposed by magnetic fields and attaining optical nonreciprocal, such as
optical nonlinearity \cite{2014NatPh..10..394P,PhysRevLett.118.033901,
2018NaPho..12..744Z,PhysRevLett.121.203602,
PhysRevLett.123.233604,PhysRevResearch.2.033517,
Tang2021BroadintensityrangeON,PhysRevLett.128.083604}, spatiotemporal
modulation of the medium\cite%
{PhysRevLett.110.093901,PhysRevLett.110.223602,article}, optomechanical
resonators\cite{articleHafezi,2016NaPho..10..657S,2016NatCo...713662R},
spinning resonators\cite%
{2018Natur.558..569M,PhysRevLett.121.153601,2018Optic...5.1424J,
PhysRevLett.128.213605,Liu230}, chiral quantum optical systems\cite%
{PhysRevX.5.041036,2014PhRvA..90d3802X,2016Sci...354.1577S,
Rauschenbeutel2016ChiralQO,2020Optic...7.1690J,2021SciA....7.8924D}, and
non-Hermitian systems\cite%
{PhysRevLett.114.253601.2015,2010NatPh...6..192R,PhysRevLett.113.123004,
PhysRevLett.110.234101,2014NatPh..10..394Peng,articleChang,2020NanoL..20.7594Z, Wang2020AllOpticalMR,Wang2012OpticalDM,Xu100}%
.

Previous research has mainly focused on the theoretical and experimental
research of nonreciprocal isolators\cite%
{PhysRevA.96.033838,XIA2019197,2013NaPho...7..579J,6678206,
PhysRevX.14.021002,PhysRevB.97.115431,PhysRevA.78.023804}. However, the
exploration of nonreciprocal photonic band gaps has been relatively limited.
In recent studies, attention has been directed towards investigating
nonreciprocal single photon band structures by exploring the chiral coupling
between one-dimensional coupled resonator optical waveguides and two-level
quantum emitter arrays\cite{PhysRevLett.128.203602}. Another approach
involves achieving nonreciprocal optical band gaps through a one-dimensional
optomechanical resonator array\cite{PhysRevA.108.063516}. Both methods
exploit the presence of two degenerate modes in the whispering-gallery-mode (WGM) resonator,
with one mode being selectively coupled with the external field to realize a
nonreciprocal single photon band gap. Inspired by these findings, we propose
using a rotating resonator chain to achieve a photonic band structure with
nonreciprocity. The Sagnac-Fizeau shift due to rotation breaks the energy degeneracy of the two modes in the WGM resonator, resulting in nonreciprocity of photon transport. By studying the different rotation modes in a single cell in the resonator chain, we show the different photonic band structures. The characteristics of photon transport are analyzed through the band structure.

The paper is organized as follows. In Sec.~\ref{Sec:2}, we introduce the Hamiltonians and the band structure of coupled rotating resonator chain. In Sec.~\ref{Sec:3}, we use Schr$\ddot{o}$dinger equation to write out the matrix of transmission probability amplitude, and report and analyze the photon transmission spectrum. The conclusions are presented in Sec.~\ref{Sec:4}

\begin{figure*}[tbp]
\includegraphics[width=0.8 \textwidth]{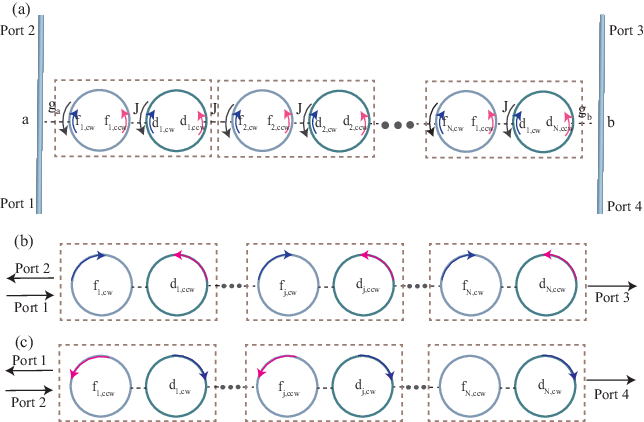}
\caption{(Color online) (a) Schematic of a 1D spinning WGM microresonators
array with $N$ unit cells and the input-output waveguides. Dashed boxes
indicate the unit cells. (b) The chain for the forward-input case. (c) The
chain for the backward-input case.}
\label{fig1}
\end{figure*}

\section{\label{Sec:2} Model and its Band structure}

Two waveguides labeled $a$ and $b$ respectively form a four port arrangement
with each port labeled with the numbers 1-4 as shown in Fig.~\ref{fig1}. A
forward-propagating (backward-propagating) photon with group velocity $v_{g}$
are created by the bosoinc field operator $\hat{a}_{F_{x}(B_{x})}^{\dagger }(%
\hat{b}_{F_{x}(B_{x})}^{\dagger })$ along the waveguide $a$ ($b$) at
position $x$. A finite-length chain with $N$ unit cells of
WGM microresonators is sandwiched between two
waveguides, each lossless resonator supports a degenerate pair of clockwise
(CW) and counterclockwise (CCW) propagating-wave modes with the frequency $%
\omega _{c}$. The resonators are divided into $f$ and $d$ sublattice groups,
the CW (CCW) mode in $f$ resonator couples to the CCW (CW) modes in $d$
resonator with strength $J$. When the WGM rotates at an angular velocity $%
\Omega _{\beta }$, the rotation-induced Sagnac-Fizeau shift
\begin{equation}
\Delta _{F\beta }=\frac{nR\Omega _{\beta }\omega _{c}}{c}\left( 1-\frac{1}{%
n^{2}}-\frac{\lambda }{n}\frac{dn}{d\lambda }\right) ,(\beta =f,d)
\end{equation}%
is introduced, where $n$ ($R$) expresses the refractive index (radius) of
the resonator and $c$ ($\lambda $) represents the speed (wavelength) of
light in vacuum, $\Omega _{\beta }>0$ ($\Omega _{\beta }<0$) indicates CW
(CCW) rotation of the resonator. The dispersion term $dn/{d\lambda }$,
characterizing the relativistic origin of the Sagnac shift, is relatively
small in typical materials ($\sim 1\%$)\cite{Grigorii}.

Light launched into port 1 or 4 (2 or 3), i.e. forward-input
(backward-input) case, drives the CW (CCW) mode in the $f$ resonator and the
CCW (CW) mode in the $d$ resonator, which is referred as $CW_{f}-CCW_{d}$ ($%
CCW_{f}-CW_{d}$) supermodes. Under the rotating wave approximation, the
total Hamiltonian $\hat{H}=\hat{H}_{1}+\hat{H}_{2}$ is the sum of two
parts,, the first part
\begin{eqnarray}
\hat{H}_{1} &=&\sum_{j=1}^{N}\omega _{f,cw}\hat{f}_{j,cw}^{\dagger }\hat{f}%
_{j,cw}+\sum_{j=1}^{N}\omega _{d,ccw}\hat{d}_{j,ccw}^{\dagger }\hat{d}%
_{j,ccw}  \notag \\
&&-iv_{g}\int dx\hat{a}_{F_{x}}^{\dagger }\frac{\partial }{\partial x}\hat{a}%
_{F_{x}}-iv_{g}\int dx\hat{b}_{F_{x}}^{\dagger }\frac{\partial }{\partial x}%
\hat{b}_{F_{x}}  \notag \\
&&+J\sum_{j=1}^{N}\left( \hat{f}_{j,cw}^{\dagger }\hat{d}_{j,ccw}+H.c.%
\right) +J\sum_{j=1}^{N-1}\left( \hat{d}_{j,ccw}^{\dagger }\hat{f}%
_{j+1,cw}+H.c.\right)   \notag \\
&&+g_{a}\left( \hat{a}_{F_{0}}^{\dagger }\hat{f}_{1,cw}+H.c.\right)
+g_{b}\left( \hat{b}_{F_{0}}^{\dagger }\hat{d}_{N,ccw}+H.c.\right),
\label{02}
\end{eqnarray}%
depicts the forward-input case and the backward-input case is described by
\begin{eqnarray}
\hat{H}_{2} &=&\sum_{j=1}^{N}\omega _{f,ccw}\hat{f}_{j,ccw}^{\dagger }\hat{f}%
_{j,ccw}+\sum_{j=1}^{N}\omega _{d,cw}\hat{d}_{j,cw}^{\dagger }\hat{d}_{j,cw}
\notag \\
&&+iv_{g}\int dx\hat{a}_{B_{x}}^{\dagger }\frac{\partial }{\partial x}\hat{a}%
_{B_{x}}+iv_{g}\int dx\hat{b}_{B_{x}}^{\dagger }\frac{\partial }{\partial x}%
\hat{b}_{B_{x}}  \notag \\
&&+J\sum_{j=1}^{N}\left( \hat{f}_{j,ccw}^{\dagger }\hat{d}%
_{j,cw}+H.c.\right) +J\sum_{j=1}^{N-1}\left( \hat{d}_{j,cw}^{\dagger }\hat{f}%
_{j+1,ccw}+H.c.\right)   \notag \\
&&+g_{a}\left( \hat{a}_{B_{0}}^{\dagger }\hat{f}_{1,ccw}+H.c.\right)
+g_{b}\left( \hat{b}_{B_{0}}^{\dagger }\hat{d}_{N,cw}+H.c.\right),
\end{eqnarray}%
where $\omega _{\beta ,cw}=\omega _{c}-\Delta _{F\beta }$ and $\omega
_{\beta ,ccw}=\omega _{c}+\Delta _{F\beta }$, $g_{a}$ ($g_{b}$) is the
coupling strength between $f_{1}$ ($d_{N}$) resonator and waveguide $a$ ($b$%
).
By applying the Fourier transform to the supermodes with the periodic
boundary condition, the dispersion relation%
\begin{equation}
\omega _{1k}^{\pm }=\omega _{c}+\frac{\Delta _{Fd}-\Delta _{Ff}}{2}\pm \sqrt{%
4J^{2}\cos ^{2}\frac{k}{2}+\frac{\left( \Delta _{Fd}+\Delta _{Ff}\right) ^{2}%
}{4}},  \notag
\end{equation}%
is obtained for the $CW_{f}-CCW_{d}$ supermode with wave number $k\in
\lbrack 0,2\pi ]$ as well as
\begin{equation}
\omega _{2k}^{\pm }=\omega _{c}-\frac{\Delta _{Fd}-\Delta _{Ff}}{2}\pm \sqrt{%
4J^{2}\cos ^{2}\frac{k}{2}+\frac{\left( \Delta _{Fd}+\Delta _{Ff}\right) ^{2}%
}{4}}.  \notag
\end{equation}%
for the $CCW_{f}-CW_{d}$ supermode. So there are two bands for $%
CW_{f}-CCW_{d}$ ($CCW_{f}-CW_{d}$) supermode: one at the frequencies between
$\omega _{10}^{+}$($\omega _{20}^{+}$) and $\omega _{1\pi }^{+}$($\omega
_{2\pi }^{+}$), the other between $\omega _{10}^{-}$($\omega _{20}^{-}$) and
$\omega _{1\pi }^{-}$($\omega _{2\pi }^{-}$). The band structure of the
total system is determined by the rotating direction and magnitude. In Figs.%
\ref{fig2}(a)-(e), we have plotted the single-photon band structures of
supermodes in different rotation directions and magnitudes. The red (blue)
line indicates that the $CW_{f}-CCW_{d}$ ($CCW_{f}-CW_{d}$) supermode is
excited. For static resonators, besides energies that excite $%
CW_{f}-CCW_{d}$ supermode and $CCW_{f}-CW_{d}$ supermode are degenerate $%
\omega _{1k}^{\pm }=\omega _{2k}^{\pm }$, the upper band (red line) touches
the lower one at $k=\pi $, see Fig.~\ref{fig2}(a). It displays a band with
size $\left\vert \omega _{10}^{+}-\omega _{1\pi }^{-}\right\vert $. The
rotation induces gaps between the bands. When all the resonators rotate in
the same direction with the same angular velocity (i.e., $\Omega _{f}=\Omega
_{d}\neq 0$), only a gap with width $2\Delta _{Ff}$ appears in the total
system and the bands of both $CW_{f}-CCW_{d}$ supermode and $CCW_{f}-CW_{d}$
supermode are degenerate, see Fig.~\ref{fig2}(b). When one of the rotating
velocities in either the $f$ sublattice or $d$ sublattice decreases (e.g., $%
\Omega _{f}>\Omega _{d}>0$), the gap of the total system becomes smaller
since the two upper bands $\omega _{1k}^{+}$ and $\omega _{2k}^{+}$ are
partially degenerate, so do the two lower bands, see Fig.~\ref{fig2}(c).
When the angular velocity in either sublattices decreases to zero, for
example, all $d$ resonators are stationary and all $f$ resonators rotate
clockwise, the gap of the total system disappears as the red solid line
touches the the blue dashed line in Fig.~\ref{fig2}(d), however, the gap
still remains for each supermode, the width of the band gap is $\Delta _{Ff}$%
, see the area between either the red lines or the blue lines in Fig.~\ref%
{fig2}(d). As $\Omega _{d}$ decreases untill $\Omega _{d}=-\Omega _{f}$,
i.e., all resonators of the $f$ sublattice rotate clockwise and all
resonators of the $d$ sublattice rotate counterclockwise, we obtain two
homogeneous coupled resonator arrays. In this case, there is no gap in the
total system and no gap in either supermodes, and the supermodes are
partially degenerate. As $\Omega _{d}$ decreases further, three bound gaps
can be observed, see Fig.~\ref{fig2}(e). In Fig.~\ref{fig2}(f), we have
plotted the variation of the energy with the Sagnac-Fizeau displacement for
one supermode, it shows that the width of the gap increases as the magnitude
of the angular velocity increase.

\section{\label{Sec:3} Nonreciprocal single-photon band gap}


In this section, we examine the transmission of finite-length resonators in
a supermode for a monochromatic photon incoming through port $i$ into port $%
j $ with amplitude denoted by $t_{i\rightarrow j}$ $(i,j={1,2,3,4})$. The
transmission probability from port $i$ to port $j$
\begin{equation}
T_{i\rightarrow j}=|t_{i\rightarrow j}|^{2}.
\end{equation}%
is defined as the square norm of its corresponding amplitude. When the $%
CW_{f}-CCW_{d}$ supermode is excited, the photon can not propagate from port
1 (4) to ports 1 and 4, i.e. $T_{1\rightarrow 1}=T_{1\rightarrow 4}=0$ ($%
T_{4\rightarrow 4}=T_{4\rightarrow 1}=0$) because the modes that photons
entering from port 1 (4) cannot be coupled to the CCW (CW) mode of the $f$ ($%
d$) resonator due to the direction of the photon momentum in the opposite
direction. So the relation $T_{1\rightarrow 2}+T_{1\rightarrow 3}=1$ ($%
T_{4\rightarrow 3}+T_{4\rightarrow 2}=1$) guarantees the probability
conservation for the incident photon. When the $CCW_{f}-CW_{d}$ is excited,
the photon can not travel from port 2 (3) to ports 2 and 3, so the
transmission probability $T_{2\rightarrow 2}=T_{2\rightarrow 3}=0$ ($%
T_{3\rightarrow 3}=T_{3\rightarrow 2}=0$).

In the single excitation subspace, the corresponding eigenvectors of
Hamiltonian $\hat{H}_{1}$ and $\hat{H}_{2}$ read
\begin{eqnarray}
\left\vert \psi _{1}\right\rangle &=&\sum_{j}C_{j,cw}\hat{f}_{j,cw}^{\dagger
}\left\vert \varnothing \right\rangle +\sum_{j}D_{j,ccw}\hat{d}%
_{j,ccw}^{\dagger }\left\vert \varnothing \right\rangle  \notag \\
&&+\int dxA_{F}\left( x\right) \hat{a}_{F_{x}}^{\dagger }\left\vert
\varnothing \right\rangle +\int dxB_{F}\left( x\right) \hat{b}%
_{F_{x}}^{\dagger }\left\vert \varnothing \right\rangle,  \notag \\
\left\vert \psi _{2}\right\rangle &=&\sum_{j}C_{j,ccw}\hat{f}%
_{j,ccw}^{\dagger }\left\vert \varnothing \right\rangle +\sum_{j}D_{j,cw}%
\hat{d}_{j,cw}^{\dagger }\left\vert \varnothing \right\rangle  \notag \\
&&+\int dxA_{B}\left( x\right) \hat{a}_{B_{x}}^{\dagger }\left\vert
\varnothing \right\rangle +\int dxB_{B}\left( x\right) \hat{b}%
_{B_{x}}^{\dagger }\left\vert \varnothing \right\rangle,  \label{04}
\end{eqnarray}%
where $C_{j,\alpha }$ and $D_{j,\alpha }$ are the probability amplitudes to
find a photon in the $\alpha $-mode of the $j$th resonator in the $f$
sublattice and $d$ sublattice, $A_{F\left( B\right) }\left( x\right) \ $and $%
B_{F\left( B\right) }\left( x\right) \ $stand for the excitation amplitudes
in waveguide $a$ and $b$ respectively, which contain just a single photon in
the forward-propagating (backward-propagating) field mode at position $x$.
Ket $\left\vert \varnothing \right\rangle $ is the ground state of the total
system. Substituting Eq.(\ref{04}) into the Schr\"{o}dinger equation $\hat{H}%
\left\vert \psi \right\rangle =E\left\vert \psi \right\rangle $, we obtain
the equation
\begin{eqnarray}
\left( \Delta +\Delta _{Ff}\right) C_{j,cw} &=&J D_{j,ccw}+g_{a}A_{F}\left(
0\right) \delta _{j,1}+J D_{j-1,ccw}\Theta \left( j-1\right),  \notag
\label{05} \\
\left( \Delta -\Delta _{Fd}\right) D_{j,ccw} &=&J C_{j,cw}+g_{b}B_{F}\left(
0\right) \delta _{j,N}+J C_{j+1,cw}\Theta \left( N-j\right),  \notag \\
\left( \Delta +\omega _{c}\right) A_{F}\left( x\right)
&=&g_{a}C_{1,cw}\delta \left( x\right) -iv_{g}\frac{\partial A_{R}\left(
x\right) }{\partial x},  \notag \\
\left( \Delta +\omega _{c}\right) B_{F}\left( x\right)
&=&g_{b}D_{N,ccw}\delta \left( x\right) -iv_{g}\frac{\partial B_{R}\left(
x\right) }{\partial x}.
\end{eqnarray}%
for the amplitudes $C_{j,cw}$, $D_{j,ccw}$, $A_{F}(x)$ and $B_{F}(x)$ in the
forward-input case, and
\begin{eqnarray}
\left( \Delta -\Delta _{Ff}\right) C_{j,ccw} &=&J D_{j,cw}+J
D_{j-1,cw}\Theta \left( j-1\right) +g_{a}A_{B}\left( 0\right) \delta _{j,1},
\notag \\
\left( \Delta +\Delta _{Fd}\right) D_{j,cw} &=&J C_{j,ccw}+J
C_{j+1,ccw}\Theta \left( N-j\right) +g_{b}B_{B}\left( 0\right) \delta _{j,N},
\notag \\
\left( \Delta +\omega _{c}\right) B_{B}\left( x\right)
&=&g_{b}D_{N,cw}\delta \left( x\right) +iv_{g}\frac{\partial B_{L}\left(
x\right) }{\partial x},  \notag \\
\left( \Delta +\omega _{c}\right) A_{B}\left( x\right)
&=&g_{a}C_{1,ccw}\delta \left( x\right) +iv_{g}\frac{\partial A_{L}\left(
x\right) }{\partial x}.
\end{eqnarray}%
for the amplitudes $C_{j,ccw}$, $D_{j,cw}$, $A_{B}(x)$ and $B_{B}(x)$ in the
backward-input case, where detuning $\Delta =E-\omega _{c}$.
\begin{figure*}[tbp]
\begin{center}
\vspace{-2cm}
\includegraphics[width=1.2 \textwidth]{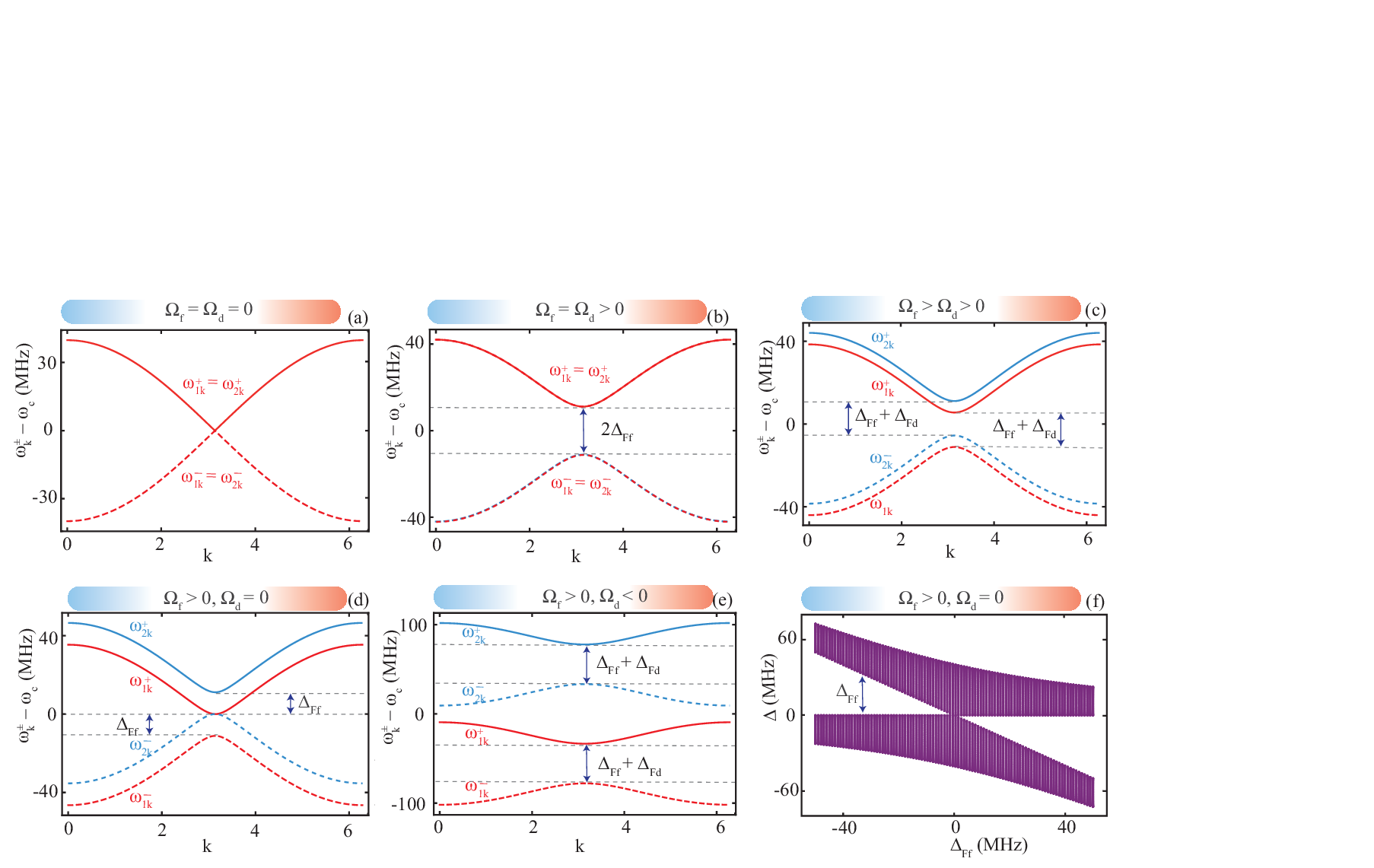}
\caption{(Color online) Single-photon band structure (a)-(e). The energy
detuning versus Sagnac-Fizeau shift $\Delta _{Ff}$, when all $d$ resonators
are stationary and all $f$ resonators rotate in (f). The red line indicates
that the $CW_{f}-CCW_{d}$ supermode is excited. The blue line indicates
that the $CCW_{f}-CW_{d}$ supermode is driven. The parameters are setting as
follow: (a) $\Omega _{f}=\Omega _{d}=0$, (b) $\Omega _{f}=\Omega _{d}=100$%
kHz,(c)$\Omega _{f}=100$ kHz, $\Omega _{d}=50$ kHz, (d)$\Omega _{f}=100$
kHz, $\Omega _{d}=0$, or $\Omega _{f}=0$, $\Omega _{d}=-100$ kHz , and (d) $%
\Omega _{f}=300$ kHz, $\Omega _{d}=-700$ kHz. Other parameters can be found
in the main text.}
\label{fig2}
\end{center}
\end{figure*}
\begin{figure*}[tbp]
\includegraphics[width=0.9 \textwidth]{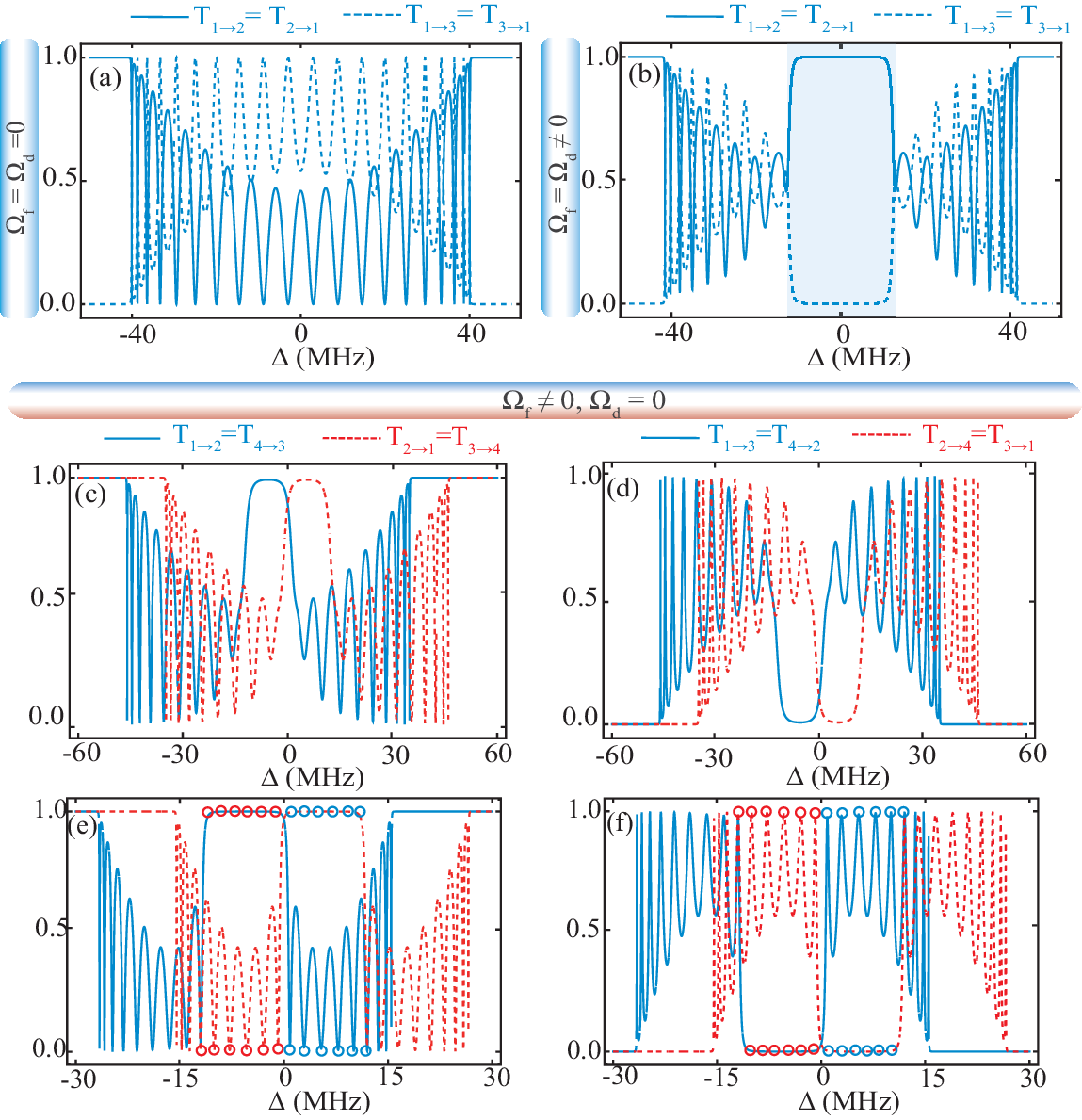}
\centering
\caption{(Color online) Single-photon transmission probability trough a
chain of $2N=20$ resonators. In (a), all resonators are stationary. In (b), all resonators rotate in the same direction and size. In (c-f), all $f$ resonators rotate and all $d$ resonator are stationary. The parameters
are setting as follow: (a) $\Omega _{f}=\Omega _{d}=0$; (b) $\Omega _{f}=\Omega _{d}=100$ kHz; (c, d) $\Omega
_{f}=100$ kHz, $\Omega _{d}=0$, (e, f) $\Omega
_{f}=100$ kHz, $\Omega _{d}=0$, $J=\omega_c/(4Q)$. Other parameters can be found in the main
text.}
\label{fig3}
\end{figure*}
For a photon incident from port $1$, the wavefunctions of waveguides $a$ and
$b$ in the real space can be expressed as
\begin{subequations}
\label{07}
\begin{eqnarray}
A_{F}\left( x\right) &=&e^{i\frac{E}{v_{g}}x}\left( \theta \left( -x\right)
+t_{1\rightarrow 2}\theta \left( x\right) \right) , \\
B_{F}\left( x\right) &=&e^{i\frac{E}{v_{g}}x}t_{1\rightarrow 3}\theta \left(
x\right) .
\end{eqnarray}%
where $\theta \left( \pm x\right) $ is a step function. Applying the ansatz
in Eq.~(\ref{07}) into Eq.~(\ref{05}), we obtain the relation of
transmission amplitudes $t_{1\rightarrow 2}$ and $t_{1\rightarrow 3}$ with
probability amplitudes $C_{j,cw}$ and $D_{j,ccw}$ of the resonators, a
system of $2\left( N+1\right) $ linear equations, which allow us to write
out its matrix form as follows
\begin{widetext}
\begin{equation}
\left(
\begin{array}{c}
-1 \\
0 \\
-\frac{g_{a}}{2} \\
0 \\
\vdots  \\
0 \\
0%
\end{array}%
\right) _{2\left( N+1\right) \times 1}=\left(
\begin{array}{ccccccc}
-1 & 0 & \frac{g_{a}}{iv_{g}} & 0 & \cdots  & 0 & 0 \\
0 & -1 & 0 & 0 & \cdots  & 0 & \frac{g_{b}}{iv_{g}} \\
\frac{g_{a}}{2} & 0 & -\Delta _{Ff}-\Delta  & J  & \cdots  & 0 & 0 \\
0 & 0 & J  & \Delta _{Fd}-\Delta  & \ddots  & 0 & 0 \\
\vdots  & \vdots  & \vdots  & \ddots  & \ddots  & \vdots  & \vdots  \\
0 & 0 & 0 & 0 & \cdots  & -\Delta _{Ff}-\Delta  & J  \\
0 & \frac{g_{b}}{2} & 0 & 0 & \cdots  & J  & \Delta _{Fd}-\Delta
\end{array}%
\right) _{2\left( N+1\right) \times 2\left( N+1\right) }\left(
\begin{array}{c}
t_{1\rightarrow 2} \\
t_{1\rightarrow 3} \\
C_{1,cw} \\
D_{1,ccw} \\
\vdots  \\
C_{N,cw} \\
D_{N,ccw}%
\end{array}%
\right) _{2\left( N+1\right) \times 1}
\label{09}
\end{equation}
\end{widetext} For light launched into port $4$, the forward-propagating
mode only interacts with the $CW_{f}-CCW_{d}$ supermode, we have trasmission
ampliudes $t_{4\rightarrow 3}=t_{1\rightarrow 2}$ and $t_{4\rightarrow
2}=t_{1\rightarrow 3}$.
For a photon incident from port $2$, it is possible to propagate freely to
port $1$ and be directed to port $4$, so the ansatz for the amplitudes in
waveguide $a$ and $b$ reads
\end{subequations}
\begin{subequations}
\label{10}
\begin{eqnarray}
A_{B}\left( x\right)  &=&e^{-i\frac{E}{v_{g}}x}\left( \theta \left( x\right)
+t_{2\rightarrow 1}\theta \left( -x\right) \right) , \\
B_{B}\left( x\right)  &=&e^{-i\frac{E}{v_{g}}x}t_{2\rightarrow 4}\theta
\left( -x\right) .
\end{eqnarray}%
\end{subequations}
The continuity of wave functions give us a system of $2\left( N+1\right) $
linear equations for amplitudes
\begin{widetext}
\begin{equation}
\left(
\begin{array}{c}
-1 \\
0 \\
-\frac{g_{a}}{2} \\
0 \\
\vdots  \\
0 \\
0%
\end{array}%
\right) _{2\left( N+1\right) \times 1}=\left(
\begin{array}{ccccccc}
-1 & 0 & \frac{g_{a}}{iv_{g}} & 0 & \cdots  & 0 & 0 \\
0 & -1 & 0 & 0 & \cdots  & 0 & \frac{g_{b}}{iv_{g}} \\
\frac{g_{a}}{2} & 0 & \Delta _{Ff}-\Delta  & J  & \cdots  & 0 & 0 \\
0 & 0 & J  & -\Delta _{Fd}-\Delta  & \ddots  & 0 & 0 \\
\vdots  & \vdots  & \vdots  & \ddots  & \ddots  & \vdots  & \vdots  \\
0 & 0 & 0 & 0 & \cdots  & \Delta _{Ff}-\Delta  & J  \\
0 & \frac{g_{b}}{2} & 0 & 0 & \cdots  & J  & -\Delta _{Fd}-\Delta
\end{array}%
\right) _{2\left( N+1\right) \times 2\left( N+1\right) }\left(
\begin{array}{c}
t_{2\rightarrow 1} \\
t_{2\rightarrow 4} \\
C_{1,ccw} \\
D_{1,cw} \\
\vdots  \\
C_{N,ccw} \\
D_{N,cw}%
\end{array}%
\right) _{2\left( N+1\right) \times 1}
\label{11}
\end{equation}
\end{widetext}in the backward case. For light launched into port $3$, it is
possible for a photon to occur in ports $1$ and $4$, we have trasmission
ampliudes $t_{3\rightarrow 4}=t_{2\rightarrow 1}$ and $t_{3\rightarrow
1}=t_{2\rightarrow 4}$ since only $CW_{d}-CCW_{f}$ supermode is involved in
the interaction.

In Fig.~\ref{fig3}, we plot the transmission probability
from port $i$ to port $j$ with $N=10$ unit cells by numerically solving
Eqs.~(\ref{09}) and (\ref{11}). In our caculations, we have selected the
experimentally feasible parameters\cite{L2017OptomechanicallyIT,
2010PhRvL.104h3901G,PhysRevLett.114.114301}: $Q=3\times 10^{7}$, $R=40$ $\mu
m$, $n=1,4$, $\lambda =1.55\times 10^{-6}$, $\omega _{c}=2\pi c/\lambda $, $%
g_{a}=g_{b}=9\omega _{c}/(5Q)$, and $J=\omega _{c}/(2Q)$. Photons coming
from port $1$ is possible for a photon to hop to the $CW_{f}-CCW_{d}$
supermode. The supermode mediates photons to be routed from one wavegiude to
the other, see Figs.~\ref{fig3}. When all the resonators are
static, there are $2N$ peaks equal to one in the transmission $%
T_{1\rightarrow 3}$ ($T_{3\rightarrow 1}$) and dips equal to zero in $%
T_{1\rightarrow 2}$ ($T_{2\rightarrow 1}$), as shown in Figs.~\ref{fig3}%
(a,b). An incoming photon whose frequency is resonant with that of a
supermode might go straight and be re-emitted after absorption, and
destructive interference between this two paths is possible. A direct
consequence of this destructive interference is that the incoming photon in
waveguide $a$ ($b$) is redirected into the other waveguide $b$ ($a$) with
maximum probability. Transmissions $T_{i\rightarrow j}=T_{j\rightarrow i}$
displayed in Figs.~\ref{fig3}(a) indicates the reciprocity of photon
transmissions, which is originated from the energy degeneracy of CW and CCW
modes in each resonator. In this case, the bands of supermodes are maximally
overlapped as shown in Fig.~\ref{fig2}(a). When the rotational angular velocities of all resonators are the same direction and magnitude ($\Omega_{d}=\Omega_{f}\neq0$),
the transmission is divided into two regions of oscillating transmission separated by a band gap, as shown in Fig.~\ref{fig3}(b). Meanwhile, when the energy of the incoming photon is inside the
band gap, the photon travels straightly along its launched waveguide, see
the transmissions in the blue shadows of Fig.~\ref{fig3}(b). Here, the energy of two supermodes is degenerate as shown in Fig.~\ref{fig2}(b) and photon transport is also reciprocal. However, when all $f$ resonators and all $d$ resonators rotate in opposite directions and their rotational angular velocities are equal ($|\Omega_{f}|=|\Omega_{d}|\neq0$), we obtain two homogeneous coupled resonator chain whose photon transport is nonreciprocal. In this case, the energy of two supermodes is not degenerate. So, breaking the energy
degeneracy of two supermodes results in the nonreciprocity.  In Fig.~\ref{fig3}(c-f), we plot the transport spectrum for all $f$ resonators rotating clockwise and all $d$ resonators static, where the blue solid line indicates that the $CW_{f}-CCW_{d}$ supermode is excited, and the red dashed line indicates that the $CCW_{f}-CW_{d}$ supermode is driven. Here, the photon transport is nonreciprocal, see transmission $T_{i\rightarrow j}\neq T_{j\rightarrow i}$ in Fig.~\ref{fig3}(c,d) in which the energy spectra of supermodes are similar to that of Fig.~\ref{fig2}(d). As the coupling strength between resonator decreases, the width of each band becomes smaller but the two band gaps remain unchanged. Figure~\ref{fig3}(e,f) is
plotted by keeping parameters same to those in Fig.~\ref{fig3}(c,d) but
decreasing the coupling strength $J$. The nonreciprocity within the band gap
becomes perfect in~Fig.~\ref{fig3}(e,f) in contrast with that in Fig.~\ref{fig3}(c,d), see the small circles in Fig.~\ref{fig3}(e,f).
 For example, the
transmission probability $T_{1\rightarrow 2}=1$ and $T_{2\rightarrow 1}\neq 0
$ at the same frequency point in Fig.~\ref{fig3}(c,d), however, $%
T_{1\rightarrow 2}=1$ and  $T_{2\rightarrow 1}=0$ in Fig.~\ref{fig3}(e,f) at
several frequency points. Each supermode can open a band gap, so there are two gaps here. Each band gap can find multiple frequency points to realize the photon circulator, and the direction of the photon cycle in the two band gaps is opposite. When a photon with energy within the band gaps is
launched into those waveguide ports (input ports 1 and 4) for which it is
supposed to excite the $CW_{f}-CCW_{d}$ supermode, it remains in its initial
waveguide, however, for the two other input ports (input ports 2 and 3), the
light couples to the $CCW_{f}-CW_{d}$ supermode and can be transferred to
the other waveguide. Overall, a photon with energy located at the small red
circles in Fig.~\ref{fig3}(c,d) form a circulator that routes light with
direction $1\rightarrow 2\rightarrow 4\rightarrow 3\rightarrow 1$. A
circulator with reversed operation direction $2\rightarrow 1\rightarrow
3\rightarrow 4\rightarrow 2$ is indicated by the small blue circles in Fig.~%
\ref{fig3}(c,d).


\section{\label{Sec:4} Conclusion}


We studied a four-port device consisting of two waveguides and a coupled-spinning-resonator chain. The rotating resonator chain has two supermodes, therefore there are two band structures. By adjusting the rotational angular velocity and direction of the resonator, we can determine the band structure. If the energy bands of two supermodes completely overlap, photon transmission is reciprocal; If some parts do not overlap, photon transmission is nonreciprocal. By adjusting the coupling strength between resonators, we can achieve a perfect nonreciprocal photonic band gap. Here, there are two perfect nonreciprocal photon band gaps. Within the band gap range, there are multiple frequency windows that can be used to demonstrate multi-frequency single-photon circulators, and the direction of the circulator is opposite in different band gaps. The coupled-spinning-resonator chain with nonreciprocal single-photon band structure can be used to design novel nonreciprocal single-photon devices, which may have various applications in quantum communication and optical sensing.


\begin{acknowledgments}
This work was supported by National Natural Science Foundation of China (Grants No.~11935006, No.~12075082, No.~12247105, No.~12064010),
the science and technology innovation Program of Hunan Province (Grants No.~2020RC4047, No.~2022RC1203), National Key R$\&$D Program of China (No.~2024YFE0102400) and the Hunan Provincial major Sci-Tech Program (Grant No.~2023ZJ1010).
\end{acknowledgments}

\bibliography{references}

\end{document}